\documentclass[12pt,nofootinbib]{article}
\pdfoutput=1
\usepackage{slashed}
\usepackage{amsmath,amssymb,graphicx,multicol} 
\usepackage{epsf,color}
\usepackage[nosort]{cite}
\usepackage{cancel}
\usepackage{framed}
\usepackage{multirow}

\usepackage{color}
\usepackage{hyperref} 
\hypersetup{linktocpage=true}
\usepackage[all]{hypcap}
\hypersetup{
    bookmarks=true,         % show bookmarks bar?
    unicode=false,          % non-Latin characters in Acrobat�s bookmarks
    pdftoolbar=true,        % show Acrobat�s toolbar?
    pdfmenubar=true,        % show Acrobat�s menu?
    pdffitwindow=false,     % window fit to page when opened
    pdfstartview={FitH},    % fits the width of the page to the window
    pdftitle={Manifestations of Warped Extra Dimension in Rare Charm Decays and Asymmetries.},    % title
    pdfauthor={Ayan Paul, Alejandro de La Puente and Ikaros I. Bigi},     % author
    pdfsubject={hep-ph},   % subject of the document
    pdfcreator={TeXShop},   % creator of the document
    pdfproducer={MacTex, PDFLaTeX}, % producer of the document
    pdfkeywords={CP Violation} {Rare Charm Decays} {Models with Warped Extra Dimension}, % list of keywords
    pdfnewwindow=true,      % links in new window
    colorlinks=true,       % false: boxed links; true: colored links
    linkcolor=blue,          % color of internal links
    citecolor=magenta,        % color of links to bibliography
    filecolor=magenta,      % color of file links
    urlcolor=cyan           % color of external links
}

\newcommand{\beq}{\begin{eqnarray}}
\newcommand{\eeq}{\end{eqnarray}}

\newcommand{\centeron}[2]{{\setbox0=\hbox{#1}\setbox1=\hbox{#2}\ifdim

\wd1>\wd0\kern.5\wd1\kern-.5\wd0\fi \copy0

\kern-.5\wd0\kern-.5\wd1\copy1\ifdim\wd0>\wd1
                                       \kern.5\wd0\kern-.5\wd1\fi}}
\newcommand{\ltap}{\>\centeron{\raise.35ex\hbox{$<$}}
                               {\lower.65ex\hbox{$\sim$}}\>}
\newcommand{\gtap}{\>\centeron{\raise.35ex\hbox{$>$}}
                               {\lower.65ex\hbox{$\sim$}}\>}

\newcommand\ZZ{\hbox{\zfont Z\kern-.4emZ}}
\font\zfont = cmss10 %scaled \magstep1

\def\beq{\begin{equation}}
\def\eeq{\end{equation}}
\def\bea{\begin{eqnarray}}
\def\eea{\end{eqnarray}}

\def\bit{\begin{itemize}}
\def\eit{\end{itemize}}

\def\l{\left}
\def\r{\right}

\def\ra{\rightarrow}

\def\baa{\begin{array}}
\def\eaa{\end{array}}

\def\simgt{\mathrel{\lower2.5pt\vbox{\lineskip=0pt\baselineskip=0pt
           \hbox{$>$}\hbox{$\sim$}}}}
\def\simlt{\mathrel{\lower2.5pt\vbox{\lineskip=0pt\baselineskip=0pt
           \hbox{$<$}\hbox{$\sim$}}}}

\begin{document}
\begin{titlepage}
\begin{flushright}
%{\tt hep-ph/yymmnn}
\end{flushright}

%\vskip.5cm

\begin{center}
{\Large \bf  
Probing  Higgs couplings with high \boldmath $p_T$ Higgs production
}
\end{center}
\vskip0.5cm

\renewcommand{\thefootnote}{\fnsymbol{footnote}}
\begin{center}
{\large Aleksandr Azatov$^{a,b}$ and Ayan Paul$^{b}$~\footnote{email:  \href{mailto:aleksandr.azatov@roma1.infn.it}{aleksandr.azatov@roma1.infn.it}, \href{mailto:ayan.paul@roma1.infn.it}{ayan.paul@roma1.infn.it}}
}
\end{center}
\renewcommand{\thefootnote}{\arabic{footnote}}

%\vspace{0.3cm}
\begin{center}
{\it $^a$Dipartimento di Fisica, Universit\`a di Roma ``La Sapienza''. \\
$^b$INFN, Sezione di Roma, I-00185 Rome, Italy.} \\
\vspace*{0.1cm}
\end{center}

\vglue 1.0truecm

\begin{abstract}
\noindent 
Possible extensions of the Standard Model predict  modifications of the Higgs couplings to gluons and to the SM top quark.  The values of these two couplings can, in general, be independent.
 We discuss a way to measure these interactions by studying the Higgs production at high $p_T$ within an effective field theory formalism. We also propose an observable $r_\pm$ with reduced theoretical errors and suggest its experimental interpretation.
\end{abstract}

\end{titlepage}

%%%%%%%%%%%%%%%%%%%%%%%%%%%%%%%%%%%%%%%%%%%%%%%%%%%%%%
%%%%%%%%%%%%%%%%%%%%%%%%%%%%%%%%%%%%%%%%%%%%%%%%%%%%%%
%%%%%%%%%%%%%%
\section{Introduction}
\begin{figure}
\begin{center}
\includegraphics[scale=0.63]{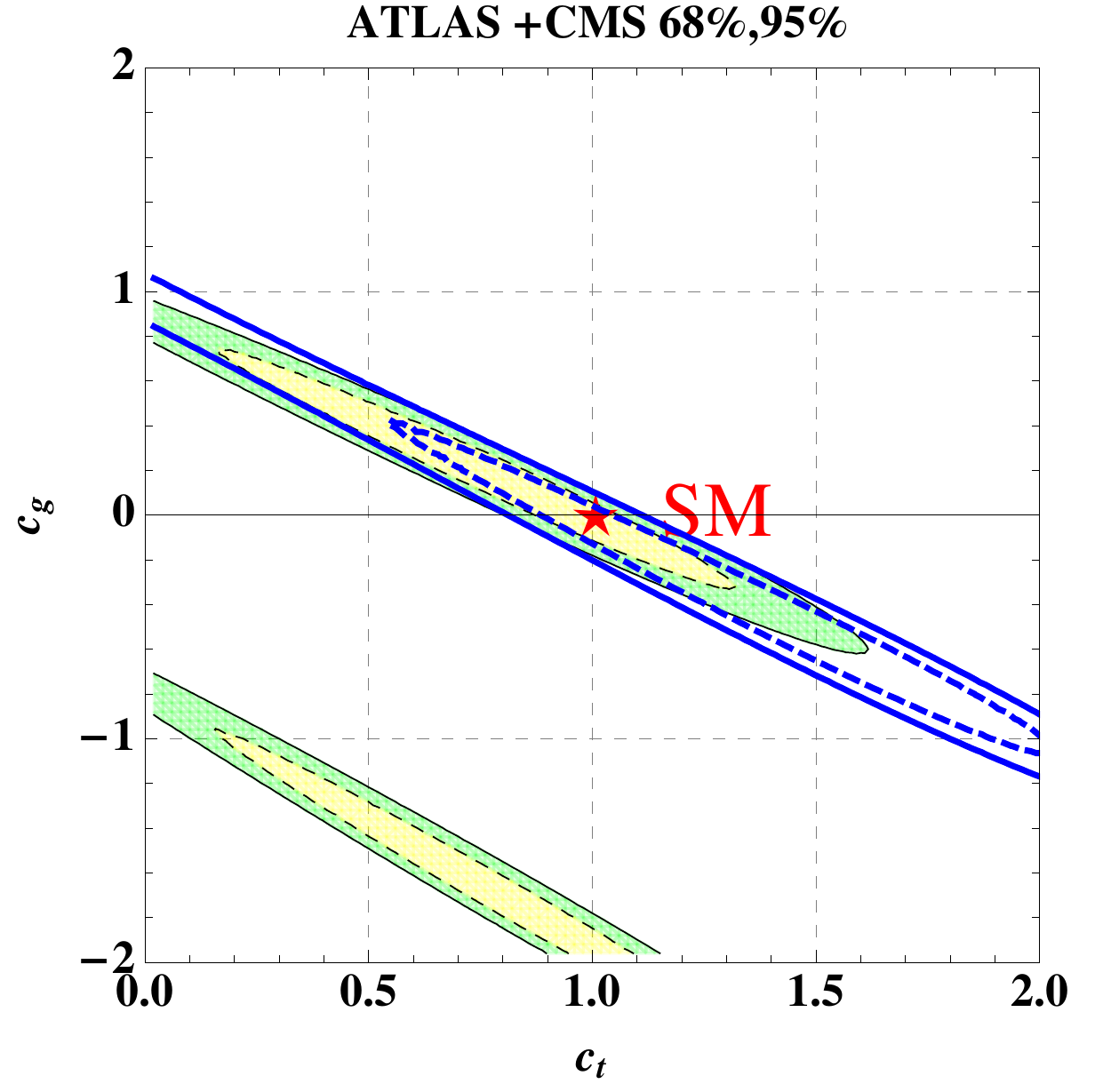}
\caption{$68\%$ and $95\%$ (yellow and green) probability contours in the $(c_t,c_g)$ plane from the Higgs couplings(based on the Lagrangian Eq.\ref{lagrang} ) The red star indicates Standard Model.
Blue lines correspond to the $68\%$ and $95\%$ contours  for the Lagrangian Eq.\ref{toplagr}.\label{ctcgfit}}
\end{center}
\end{figure}

%%%%%%%%%%%%%%
LHC has recently reported the discovery of a Higgs boson\cite{Aad:2012tfa,Chatrchyan:2012ufa}.  The properties of this newly found particle, so far, strongly resemble the properties of the Standard Model (SM) Higgs\cite{ATLASc,CMSc}. However, the task of fully establishing the nature of the electroweak symmetry breaking is far from completion. One of the ways to test the properties of the newly discovered field is to compare its couplings to the SM predictions. The current data shows agreement between theory and experiments of the order of $20\%-30\%$\cite{ATLASc,CMSc}.  It should be noted that the constraints on the values of the Higgs top Yukawa coupling come mainly from the gluon fusion measurements and the constraints from the associated production of Higgs with a top pair are still weak \cite{CMStth,CMSml,Chatrchyan:2013yea}. The discrepancies between Higgs top Yukawa coupling and the gluon fusion rate can easily arise in theories beyond the Standard Model, where the scale of the electroweak symmetry breaking  is natural. Indeed, the majority of these models predict new states, which are charged under the SU(3) colour gauge group and interact with the Higgs boson. It is  plausible that these states are beyond the direct production reach at LHC energies,  however their indirect effects can show up in the  
modifications of the Higgs couplings.
At low energies, modifications of the Higgs couplings to gluons and top quarks can be parametrized as
\bea
{\cal L}=-c_t \frac{m_t}{v} \bar t t h+\frac{g_s^2}{48\pi^2} c_g \frac{h}{v} G_{\mu\nu}G^{\mu\nu}
\label{lagrang}
,\eea
 where the ($c_t=1,c_g=0$) point corresponds to the SM. We have normalized the Higgs interaction with  gluons  in a way  that $c_g=1$  corresponds to the operator generated by an infinitely heavy top quark.   Single Higgs production occurs at the scale $m_H<m_t$ so that the Higgs Low Energy Theorems\cite{Ellis:1975ap,Shifman:1979eb} are applicable  and the effective operator controlling the Higgs couplings to gluons will be given by
\bea
O_g(m_H)\approx \frac{g_s^2}{48\pi^2} (c_g+c_t) \frac{h}{v} G_{\mu\nu}G^{\mu\nu}.
\label{maineq}
\eea
Eq. \ref{maineq} shows that  the overall gluon fusion rate will be proportional to
 $|c_g+c_t|$ and there is no way  to disentangle this combinations of the Higgs couplings from the gluon  fusion process only. The current fit of these couplings is given in Fig. \ref{ctcgfit}.
We can see that the current data is peaked along the 
$c_t+c_g=const$ line.  The black contours surrounding the yellow and green areas indicate $ 68\%$ and $95\%$ percent probability regions respectively. The fit was obtained using the likelihood from \cite{Azatov:2012qz}\footnote{We have also included a recent CMS analysis \cite{CMSml} with multiplepton final state.} and assuming that the only deviations of the Higgs couplings are the ones in the Lagrangian in Eq. \ref{lagrang}.  We can see that there is a strong degeneracy in probability contours along the $c_t+c_g= const$ direction. The only channels that break this degeneracy are $h\ra \gamma\gamma$ and $pp\ra t \bar{t} h \ra t \bar t (b\bar b,\gamma\gamma,WW,ZZ,\tau\tau) $. 
However  the discriminating power of the $h\ra \gamma \gamma$ channel is weak because the Higgs interaction with two photons is dominated by the $W$ loop. The blue dotted and dashed dotted lines on the Fig.\ref{ctcgfit} represent $ 68\%$ and $95\%$ percent probability regions within the assumption that $O_g$ operator is generated by the fields, which have the same quantum numbers as SM top quarks,  i.e.,   the effective Lagrangian is given by
\bea
{\cal L}=-c_t \frac{m_t}{v} \bar t t h+\frac{g_s^2}{48\pi^2} c_g \frac{h}{v} G_{\mu\nu}G^{\mu\nu}+\frac{e^2}{18\pi^2} c_g \frac{h}{v} \gamma_{\mu\nu}\gamma^{\mu\nu}.
\label{toplagr}
\eea
The $c_t,c_g$ degeneracy becomes even stronger in this case  since the  only channels that break it are \footnote{For the latest studies on the measurements of the top Yukawa couplings from the $pp\ra th, pp\ra t \bar{t} h $ processes see \cite{tthcoupling}}
\bea
pp\ra t \bar{t} h \ra t \bar{t} (b\bar{b},  \gamma\gamma, WW,ZZ,\bar\tau\tau).
\eea
 The second ellipse solution with negative $c_g$ is gone due to the strong constraints on the overall $h\ra\gamma\gamma$  rate, also it is interesting to note that the maximum of the probability is shifted towards values of $c_t$ greater than one due to the recent measurements of the  $\bar{t}t h$ interaction in the multilepton final sate \cite{CMSml}.
In this paper we propose to look at the $pp\ra h+j$ process as another way to resolve this degeneracy\footnote{Recently the same process was proposed for the studies of the dimension 7 operators for the Higgs gluon interactions\cite{Harlander:2013oja}. }. 
Indeed, at high transverse momentum $(p_T)$, when the center of mass energy becomes larger than the top mass $(m_t)$  and we cannot integrate it out, the cross section becomes proportional to 
 \bea
&&\frac{d\sigma}{d p_T}=\sum_i \kappa_i|f^i(p_T )c_t+c_g|^2,\\
&&\l(\frac{d\sigma^{SM}(m_t)}{d p_T}\r)/
\l(\frac{d\sigma^{SM}(m_t\ra \infty )}{d p_T}\r)=\frac{\sum_i \kappa_i f^i(p_T)^2}{\sum_i \kappa_i},
\eea
where the sum is over all non-interfering processes contributing to the $ p p\ra h+j$. For example, the recent next to leading order (NLO) calculation predicts \cite{Grazzini:2013mca}
\bea
\l(\frac{d\sigma^{SM}(m_t)}{d p_T}\r)/
\l(\frac{d\sigma^{SM}(m_t\ra \infty )}{d p_T}\r)|_{p_T=300{\rm GeV}}\sim 0.7.
\eea
Hence the measurements of the Higgs production at high $p_T$ can shed light on the $c_t$ and $c_g$ couplings.

Lastly, we would like to comment that the $(c_t, c_g)$ degeneracy is manifest in the Composite Higgs models\cite{compoH}, where the $c_t+c_g$ combination of the Higgs couplings constants was shown to be independent of the details of the spectrum of composite resonances\cite{gghc} \footnote{While this work was at the stage of completion, a similar proposal to use $h+j$ at high $p_T$ to  resolve the $(c_t,c_g)$ degeneracy and extract information about the spectrum of composite resonances was suggested  in the context of the Composite Higgs and Little Higgs scenarios \cite{Banfi:2013yoa}.}.

%%%%%%%%%%%%%%
\section{Cross section as a function of \boldmath $(c_t,c_g)$}
\begin{figure}
\begin{center}
\includegraphics[scale=0.42]{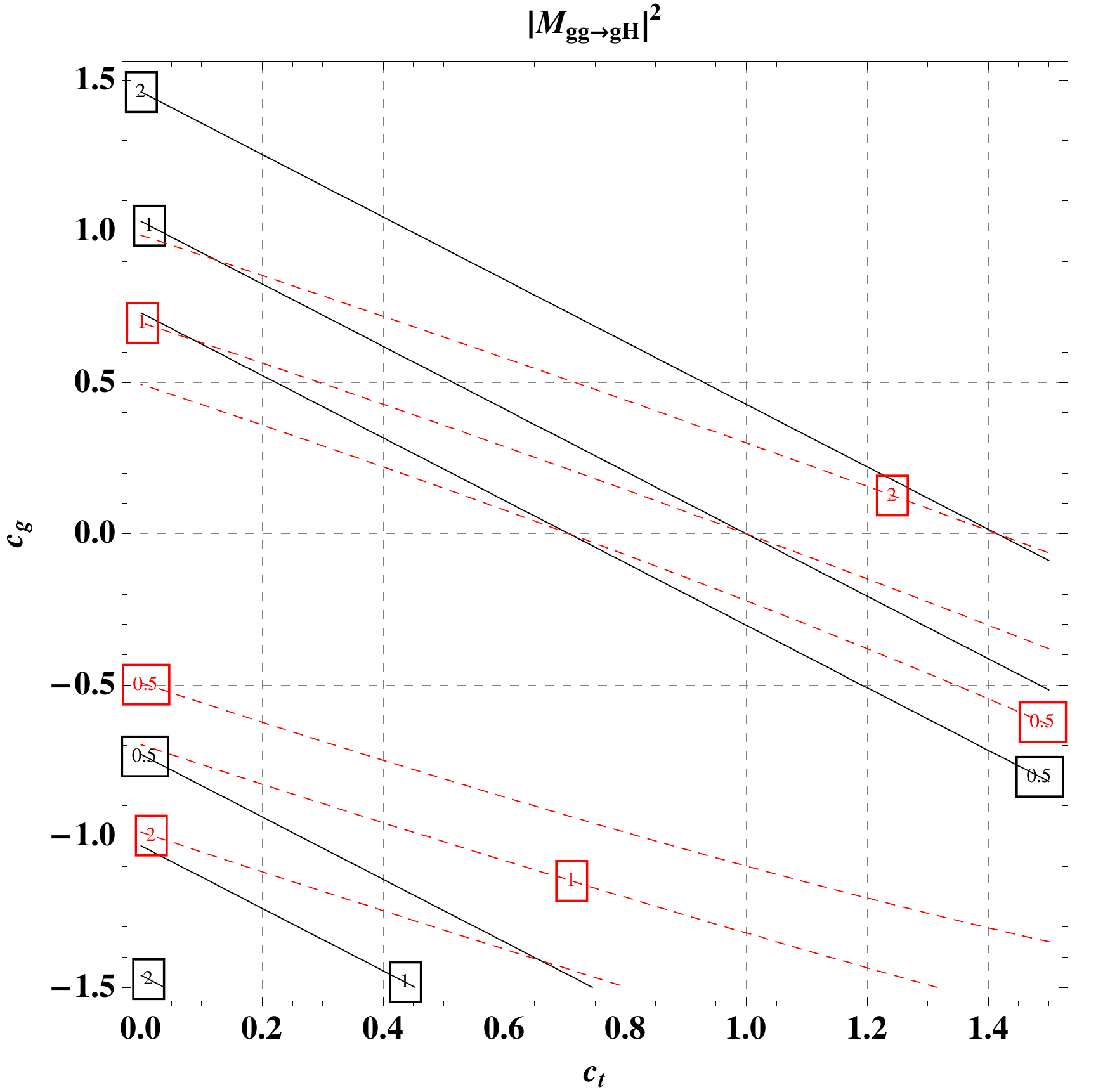}
\includegraphics[scale=0.45]{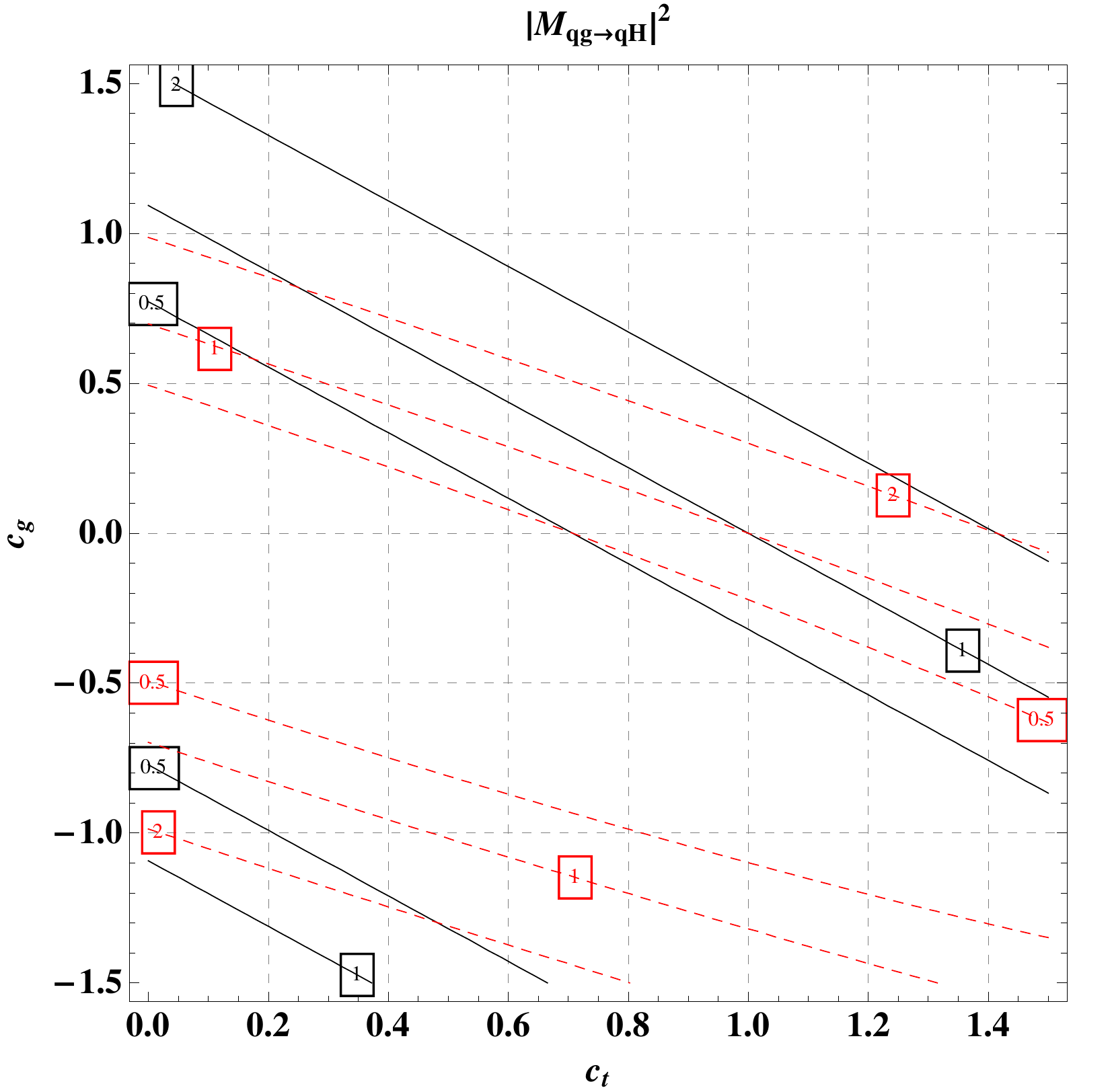}
\includegraphics[scale=0.5]{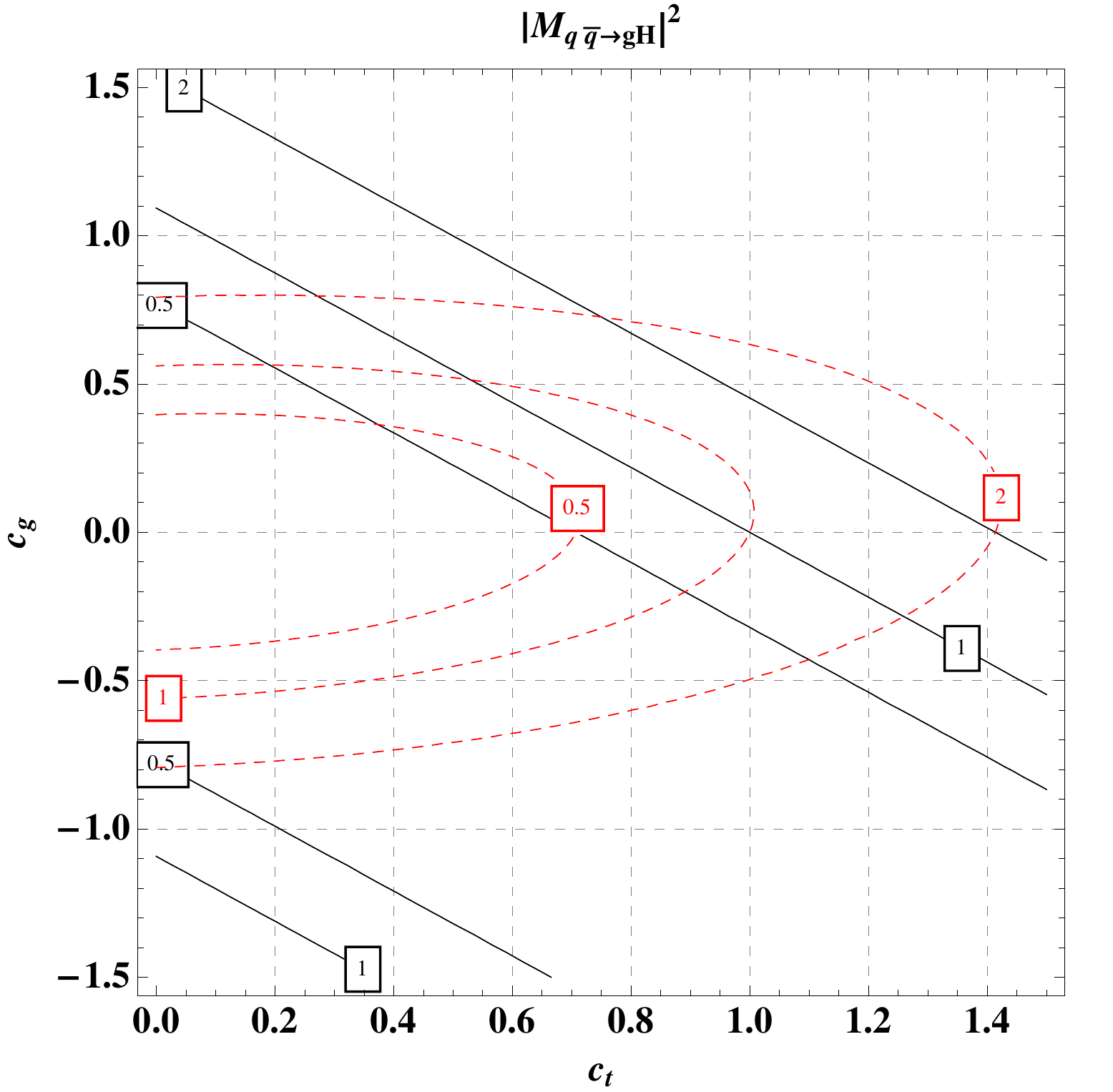}
\caption{Isocontours of $|M_i|^2$ for various values of $\hat{s}$, $\hat{t}=\hat{u}=(m_H^2-\hat {s})/2$. The red dashed line corresponds to the $\sqrt{\hat{s}}=1000$ GeV and the black solid line to the $\sqrt{\hat s}=130$ GeV. 
\label{m2} Contour labels indicate modification of the $|M_i|^2$  compared to the SM expectations.}
\end{center}
\end{figure}

%%%%%%%%%%%%%%
The dominant processes contributing to the $pp\ra h+j$ at the parton level  are $gg\ra g h,q g\ra q h,\bar{q} g\ra \bar{q} h$. The contribution from $q\bar{q}\ra h+g$ is subleading by orders of magnitude. At the leading order (LO) the matrix elements for the loops of the top quarks  were calculated by \cite{Ellis:1987xu,Baur:1989cm}. These can be easily recasted into the $(c_t,c_g)$ plane using the relation
\bea
M_i(c_t,c_g)=c_t M_i(m_t)+c_g M_i(m_t\ra \infty).
\eea
We used LoopTools \cite{Hahn:1998yk} to compute the Passarino-Veltman loop functions appearing in the  matrix elements $M_i(m_t)$. The isocontours of $|M_i|^2$ are shown in Fig. \ref{m2}.
We can see that at high collision energy the shapes of the isocontours are changed. Note that for $M_{q\bar{q}\ra gh}$ the isocontours are no more the straight lines because there is a large imaginary part corresponding to the real top pair production with the gluon in the S channel.   

Before proceeding further we would like to comment on the validity of the effective field theory (EFT) approach for parametrization of the new physics (NP) contribution in terms of the Lagrangian in  Eq.\ref{lagrang}.  We have checked numerically that, when $O_g$ is generated by the loops of new fermions, effective field theory still provides a good description of the physics if the energy of the collision is below 
\bea
\sqrt{s}\lesssim {\cal O}(M_{NF}),
\eea
where $M_{NF}$ is the mass of the new fermion\footnote{The exact upper bound on $\sqrt{s}$ depends on the contributing subprocess and for
 $gg\ra gh, (gq\ra q h, q\bar{q}\ra g h)$ is equal to  $\sqrt{s}\lesssim1.7,(1, 0.7 )M_{NF}$ respectively.  }  and we can use this inequation as an estimate of the EFT validity range.

\begin{figure}
\begin{center}
\includegraphics[scale=0.5]{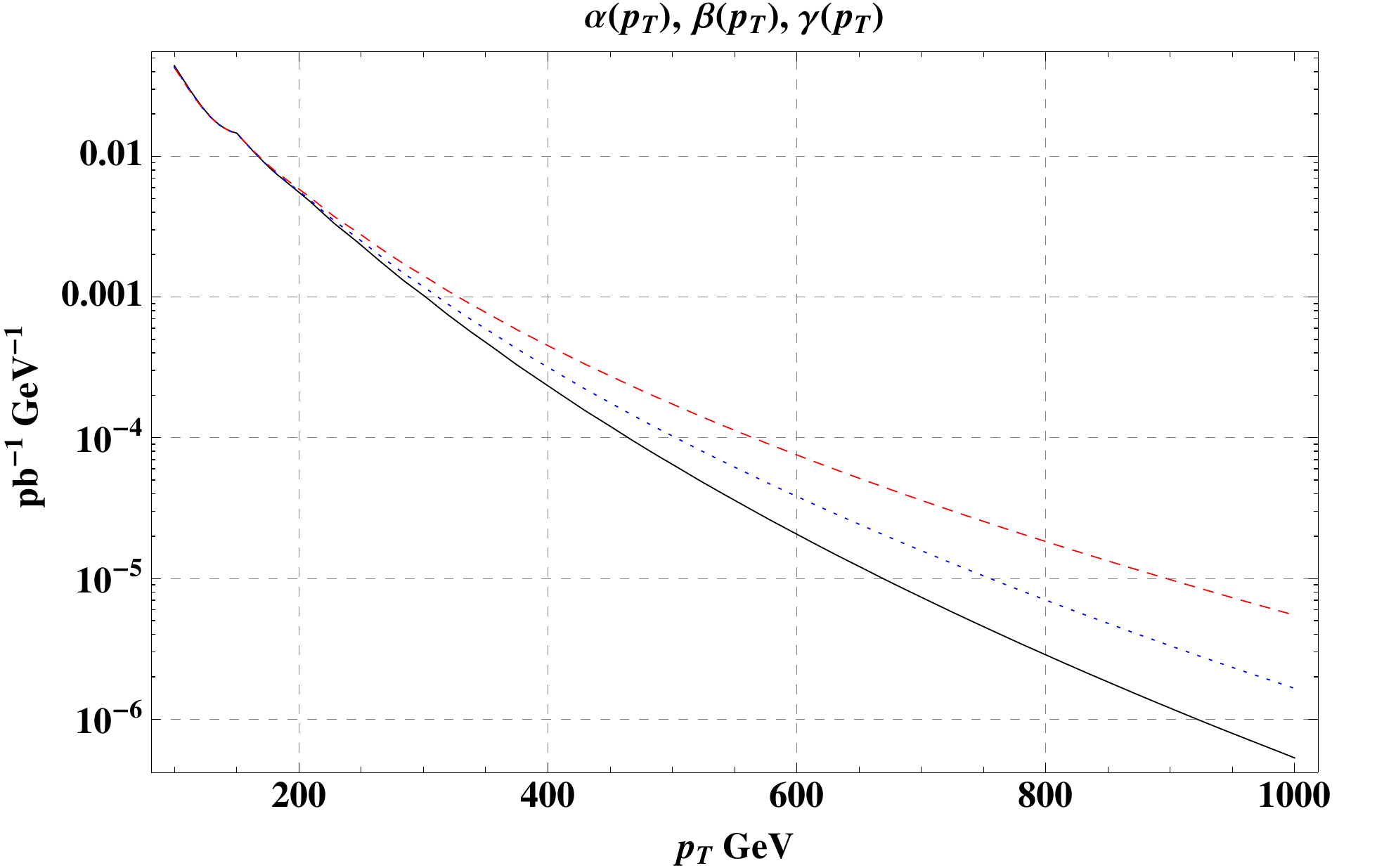}
\caption{Coefficients $\alpha,\beta,\gamma$ as a functions of $p_T$. Black (solid) -- $\alpha(p_T)$, blue (dotted) -- $\gamma(p_T)$, red (dashed) -- $\beta(p_T)$\label{figurexs}, for the center of mass energy $\sqrt{S}=14$ TeV.}
\end{center}
\end{figure}

\begin{figure}
\begin{center}
\includegraphics[width=6.9cm]{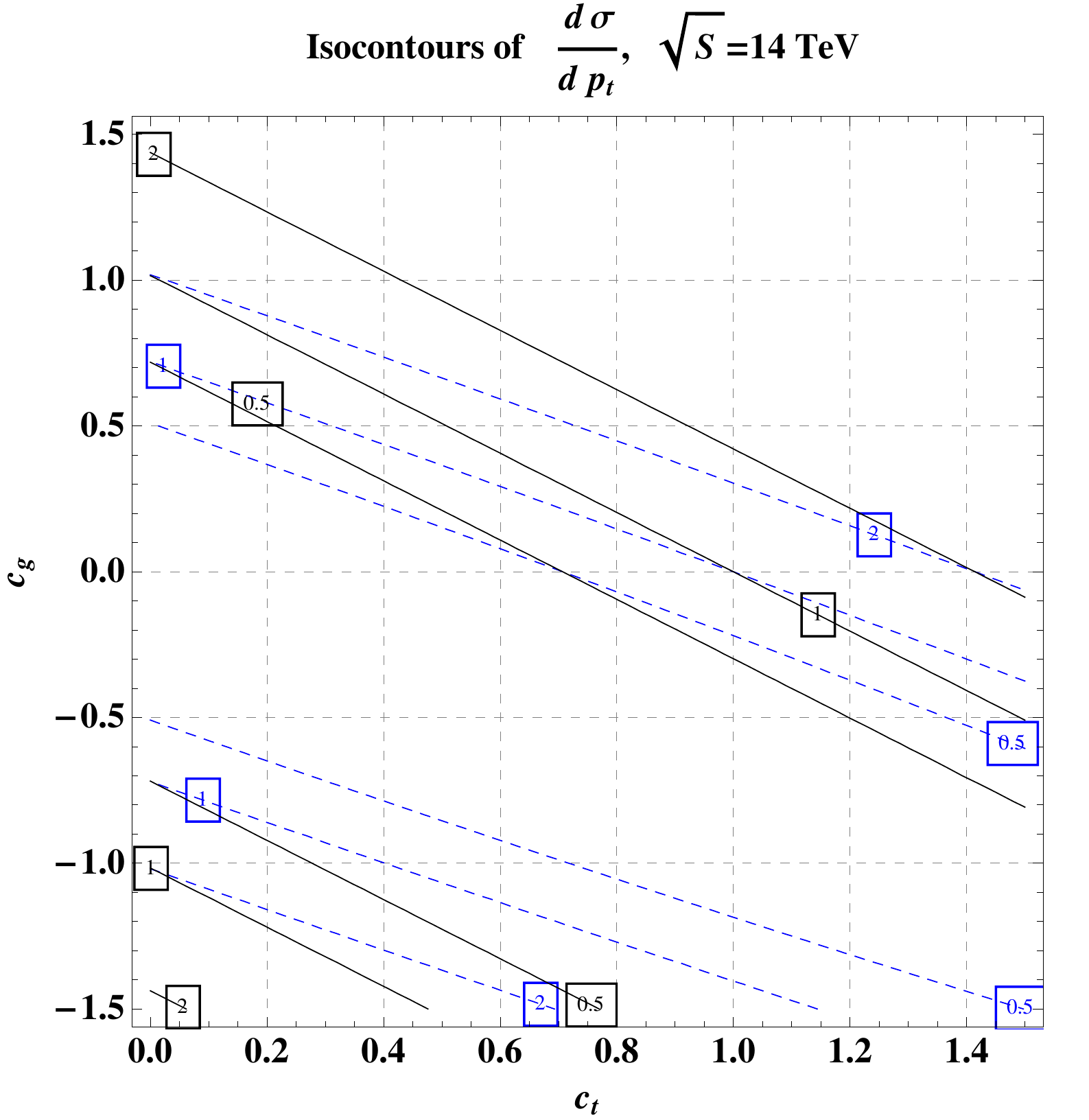}
\includegraphics[width=6.9cm]{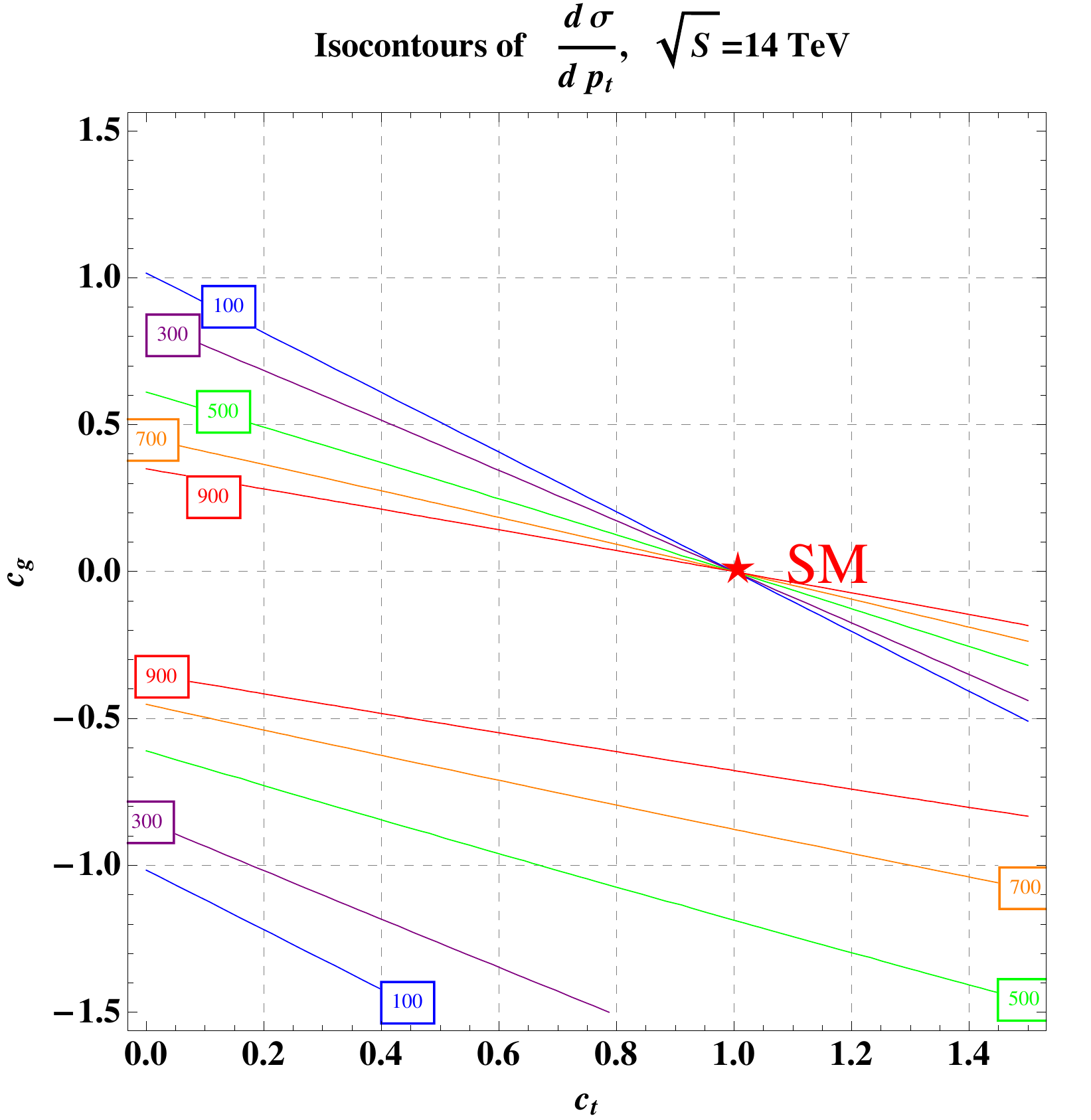}
\caption{Left -- Isocontours of $\frac{d\sigma}{d p_T}$  in the units of the SM differential cross section. Blue (dashed) --  $p_T=400$GeV, black (solid) -- $p_T=100$ GeV, SM corresponds to the (1,0) point in the plane. Right -- isocontours of the SM cross section for various $p_T$ in GeV(indicated by the labels).\label{isosimple}}
\end{center}
\end{figure}

 The partonic cross section is convoluted into the hadronic one using the procedure described in \cite{Brock:1993sz}\footnote{In this regards the works \cite{Glosser:2002gm} and \cite{Field:2003yy} prove very useful too.}, and using the PDF sets provided by MSTW2008 \cite{Martin:2009iq}.
\bea
d \sigma ( p p\ra hj)&=&\sum_{a b} f(x_1) f(x_2)d x_1 d x_2\sum \l|M(a b \ra h+ j)\r|^2 (2\pi)^4\times\nonumber\\
&&\delta^4(p_1+p_2-p_H -p_j)\frac{d^3 p_H}{2(2\pi)^3 E_H})\frac{d^3 p_j}{2(2\pi)^3 E_j}.
\eea
The LO overall cross section is  a second order polynomial  of the coefficients
$c_t$ and $c_g$
\bea
\frac{d\sigma}{d p_T}=\alpha(p_T) c_t^2+\beta(p_T) c_g^2+2\gamma(p_T) c_t c_g.
\eea
We present the values of  the coefficients $\alpha(p_T)$, $\beta(p_T)$ and $\gamma(p_T)$ for the various values of $p_T$ in Fig \ref{figurexs}.
As expected, the  difference between these coefficients grows with $p_T$.
During our calculation we have set the renormalization and the factorization scales at
\bea
\mu_r=\mu_f=\sqrt{p_T^2+m_H^2}.
\eea
To take into account higher order NLO QCD corrections we have calculated the $K(p_T)$ factors using $HqT$\footnote{While $HqT$ provides the LO+NLL and NLO+NNLL estimates, the LO estimates are not implemented with the Passarino-Veltman loop functions and hence break down at high $p_T$.} \cite{deFlorian:2011xf}, and for this choice of the renormalization and factorization scale $K$ factors are roughly $p_T$ independent and are approximately$K(p_T)\sim 2$. We have also verified that the resummation effects are negligible in the $p_T$ range we have considered. 
Isocontours of the differential cross section in the $(c_t,c_g)$  plane are shown in Fig \ref{isosimple}, we can see that they strongly resemble the isocurves of the matrix elements, since the subprocess $q\bar{q}\ra g H$ is strongly subleading.

Measurement  of the cross section at fixed  $p_T$ will constrain the plane to a line (band). Combining measurements at various $p_T$ (intersection of the bands) will fix the Higgs coupling parameters $c_t$ and $c_g$.  Ideally, the whole $p_T$ distribution of all the events should be used to reconstruct the $c_t,c_g$ coefficients of the effective Lagrangian. However, to simplify the analysis and to estimate the LHC potential for $c_t,c_g$ measurements in the $pp\ra h+j$ analysis, we can categorize all the events into two bins with high and low $p_T$. 
\bea
\sigma^-(p_T<P_T)=\int_{p_T^{min}}^{P_T}\frac{d\sigma}{d p_T } d p_T, ~~~N^- =\sigma^- \times \hbox{Luminosity},\nonumber\\
\sigma^+ (p_T>P_T)=\int_{P_T}^{p_T^{max}}\frac{d\sigma}{d p_T } d p_T,~~~N^+=\sigma^+ \times \hbox{Luminosity},
\eea
where $N^\pm$ are number of events seen in the respective bins.
Calculating the real SM background is  beyond the scope of this paper, so to roughly estimate the LHC potential at $14$ TeV at very high luminosity we decided to look at the Higgs decays into the four lepton final mode $h\ra ZZ^*\ra l^-l^-l^+ l^+$ and estimated the background at the partonic level using \cite{Alwall:2011uj}.
\begin{figure}
\begin{center}
\includegraphics[width=5.7cm]{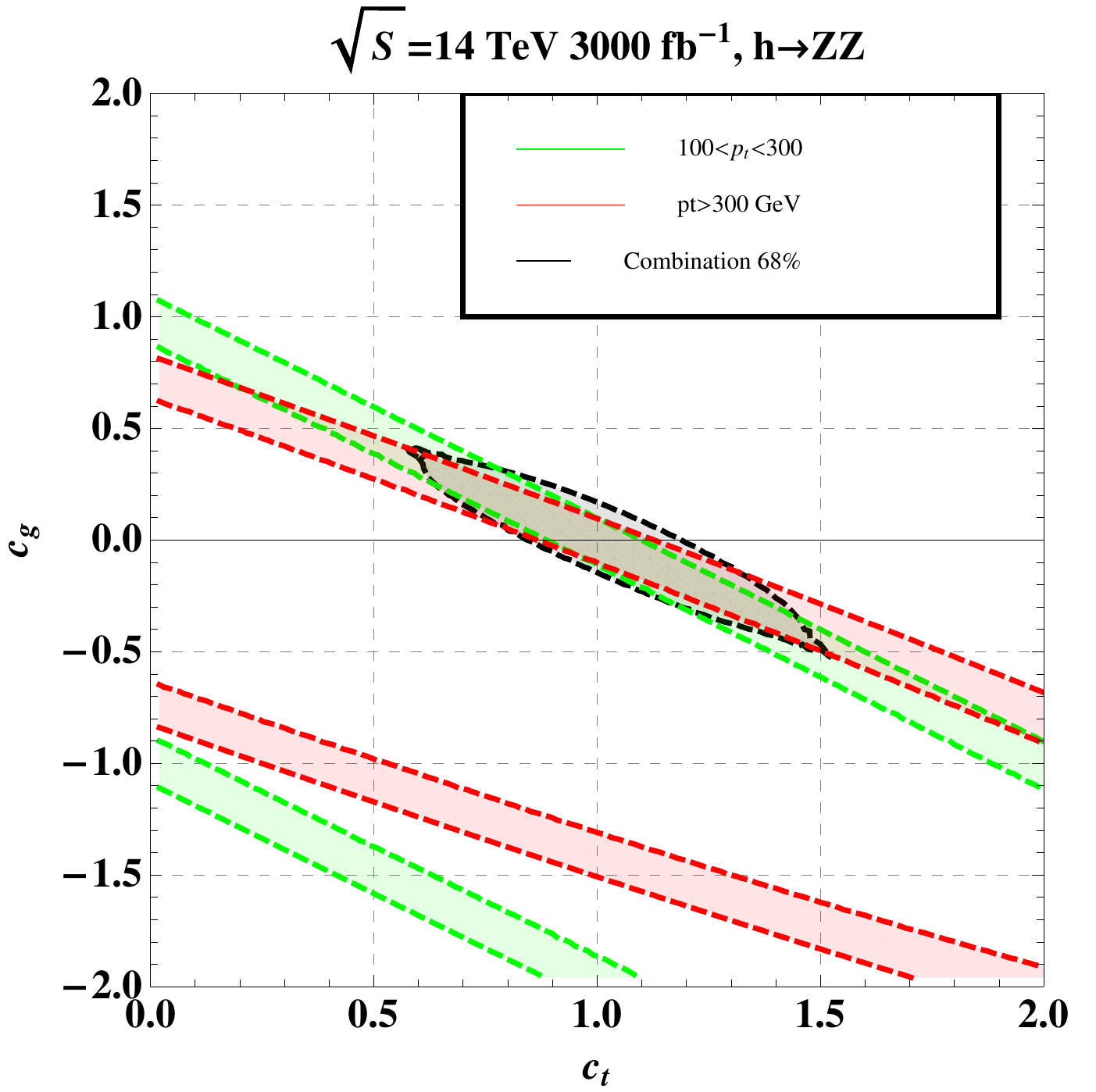}
\includegraphics[width=5.7cm]{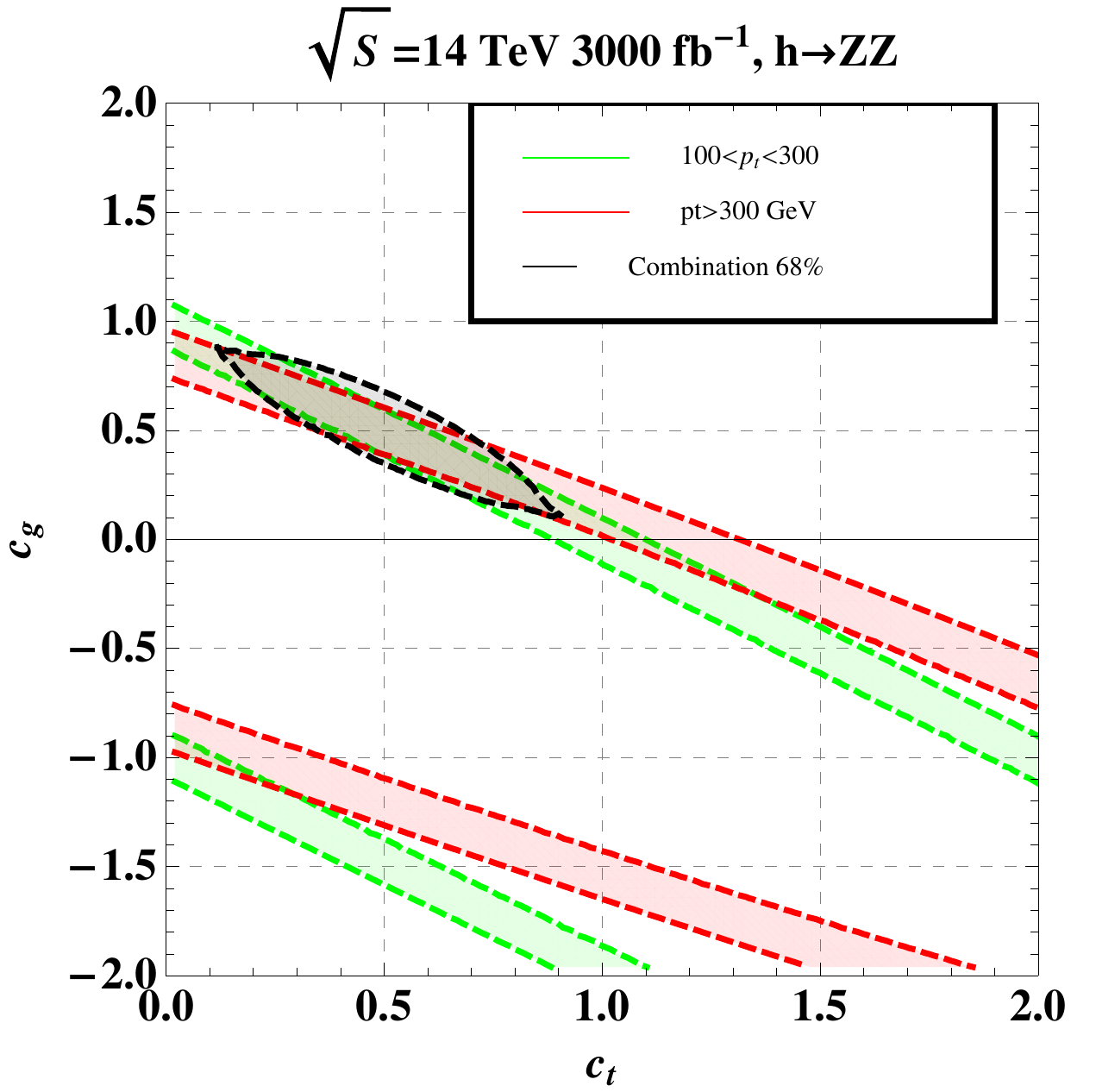}
\caption{\label{convol}
Green -- $68\%$ band coming from the $N^-$ measurement, red -- $68\%$ band coming from the $N^+$ measurement for $P_T=300$ GeV. Black is a combination assuming $100\%$ correlation between theoretical errors. The probability contours are obtained in Bayesian analysis assuming zero background for $3000 fb^{-1}$. We can see that we need very high luminosity to overcome statistical uncertainty. Left plot corresponds to the SM signal $(c_t=1,c_g=0)$, right plot to $(c_t=0.5,c_g=0.5).$}
\end{center}
\end{figure}
We assume Bayesian statistics  and treat the scale and PDF uncertainties as systematic errors, which are $100\%$ correlated for the $N^+$ and $N^- $ measurements. The results are shown in the Fig. \ref{convol}, where the plots are shown for the discriminating momenta $P_T=300$ GeV.
The green band corresponds to the  $68\%$ probability contour of the $N^-$ observable. We have treated the choice of renormalization scale and the uncertainty in the PDF as theoretical errors and varied the expected signal within this range with a Gaussian prior.  The red contour is a similar band for the $N^+$ observable. The black contour is a  combination assuming hundred percent correlation of the systematic errors which are, in our case, the choice of the renormalization/factorization scale and PDF.   Due to this correlation the overall combination contour is not just a simple overlap of the $N^\pm$ contours. However we are dominated by the low $p_T$ measurements, because the statistical uncertainty is much smaller there.  We have chosen the two categories to be discriminated by the $P_T=300$ GeV to have larger number of events in the high $p_T$ category, even though the separation between $N^+$ and $N^-$ isocurves is small. With $3000fb^{-1}$ date  we have $N^-\sim 60$ events for the SM point.

%%%%%%%%%%%%%%
\section{Understanding theory uncertainties}
%%%%%%%%%%%%%%
The combined theoretical error in the LO estimate is approximately $50\%$ and that at NLO is approximately $25\%$. Theory uncertainties come from three sources.
\begin{description}
\item{\bf Choice of renormalization and factorization scale:} \\There are different prescriptions for the choice these scales. The one which is used more prevalently for low $p_T$ analyses is proportional to the Higgs mass, $\mu_r=\mu_f=xm_H$. The other choice is proportional to the transverse energy of the Higgs,  $\mu_r=\mu_f=x\sqrt{m_H^2+p_T^2}$ with $x$ being varied between 0.5 and 2 in general. This is more relevant for high $p_T$ as it is better motivated as the ``scale'' of the physics process, hence we use it in our work. The other prescription mentioned in the literature \cite{Banfi:2013yoa} is $\mu_r=\mu_f=\frac{x}{2}\left[p_T+\sqrt{m_H^2+p_T^2}\right]$ which reduces to the latter prescription in the $p_T$ region away from $m_H$. For the LO cross section, the error from the variation of the scale leads to an error of the same order as the cross section itself. However, at the NLO, this error drops to about $25\%$. We have checked this using $HqT$ in the infinite top mass limit.

\item{\bf PDF errors:}\\ The PDF errors are of the order of $5\%$. We used the $68\%$ C.L. sets in the MSTW2008 PDF grids to determine this. Considering the scale error, this error is sub-dominant.

\item{\bf \boldmath $K$ factor:}\\ Since the NLO calculations have been performed only for an effective infinitely heavy top mass contribution, the $K$ factor for the finite top mass contribution needs to be estimated from the former. Although finite mass effects in the $K$ factor can be expected to be not very large, commenting on the significance of the same is beyond the scope of this work. The variation of the $K$ factor with the choice of scale is of $O(10\%)$, which we have numerically checked using $HqT$ in the infinite top mass limit.
\end{description}

The determination of the $c_t,c_g$ suffers from the systematic errors due to these uncertainties in the theoretical calculation. 
The scale dependence of the integrated cross sections $\sigma^+$ and $\sigma^-$ comes from the renormalization scale dependence of the strong coupling constant and the factorizations scale dependence of the PDFs. It is quite clear that the former is multiplicative and will partially drop out in ratios of any differential or partial cross sections. The factorization scale dependence is not so trivial as it comes from a convolution of the PDFs with the partonic amplitudes. However, as long as two partonic contributions are not widely different functionally and numerically, the dependence is approximately multiplicative and can be expected to bring about weak scale dependence in ratios of cross sections. For example, lets look at the ratios
\begin{equation}
R_+=\frac{\sigma^+}{\sigma^+_{SM}}\;\;{\rm and}\;\;R_-=\frac{\sigma^-}{\sigma^-_{SM}},
\end{equation}
where $\sigma^\pm_{SM}$ is defined by setting $c_t=1$ and $c_g=0$. In the absence of new physics contributions both these ratios are equal to unity. We have scanned the values of the theoretical errors in the $(c_t,c_g)$ plane and we have found out that the error on the $R_+,R_-$ is always less than $2\%$, which primarily comes from the convolution of the pdfs in the different $p_t$ regions.
 It is clear that in both the low $p_T$ (100 - 300 GeV) and the high $p_T$ (300 - 1000 GeV) regions, these ratios are almost independent of the choice of renormalization and factorization scales within the range of variation of the latter that we have chosen and almost independent of PDF errors. This statement is true regardless of what prescription we set for the choice of the scales, i.e., whether we chose fixed scales or running scales. This has a very important implication in the light of the theoretical uncertainties that shroud the calculation of cross sections in this channel. The approximate independence of the ratio from scale choice along with approximate scale (and $p_T$) independence of the $K$ factor, which are also blind to SM vs. NP contributions, implies that the ratios $R^+$ and $R^-$ are more or less independent of the order of the theoretical computation and the choice of scales. These ratios can be easily calculated at LO and will be the same even when higher order terms are added to the cross section. Note that  both $R_+$ and $R_-$ are theoretical constructs and do have any  implementations as experimental observables.

%%%%%%%%%%%%%%
\subsection{Defining and interpreting \boldmath $r_\pm$}
%%%%%%%%%%%%%%

We define the ratio
\begin{equation}
r_\pm=\frac{R_+}{R_-}.
\end{equation}
In the absence of NP contributions $r_\pm=1$. Even in the presence of NP contribution $r_\pm$ can be equal to unity if both low $p_T$ and high $p_T$ amplitudes are equivalently enhanced or diminished by NP. However, $r_\pm\neq1$ is a sure sign of the presence of a new degree of freedom and hence can be used as a discriminant from SM.

\begin{figure}
\begin{center}
\includegraphics[scale=0.55]{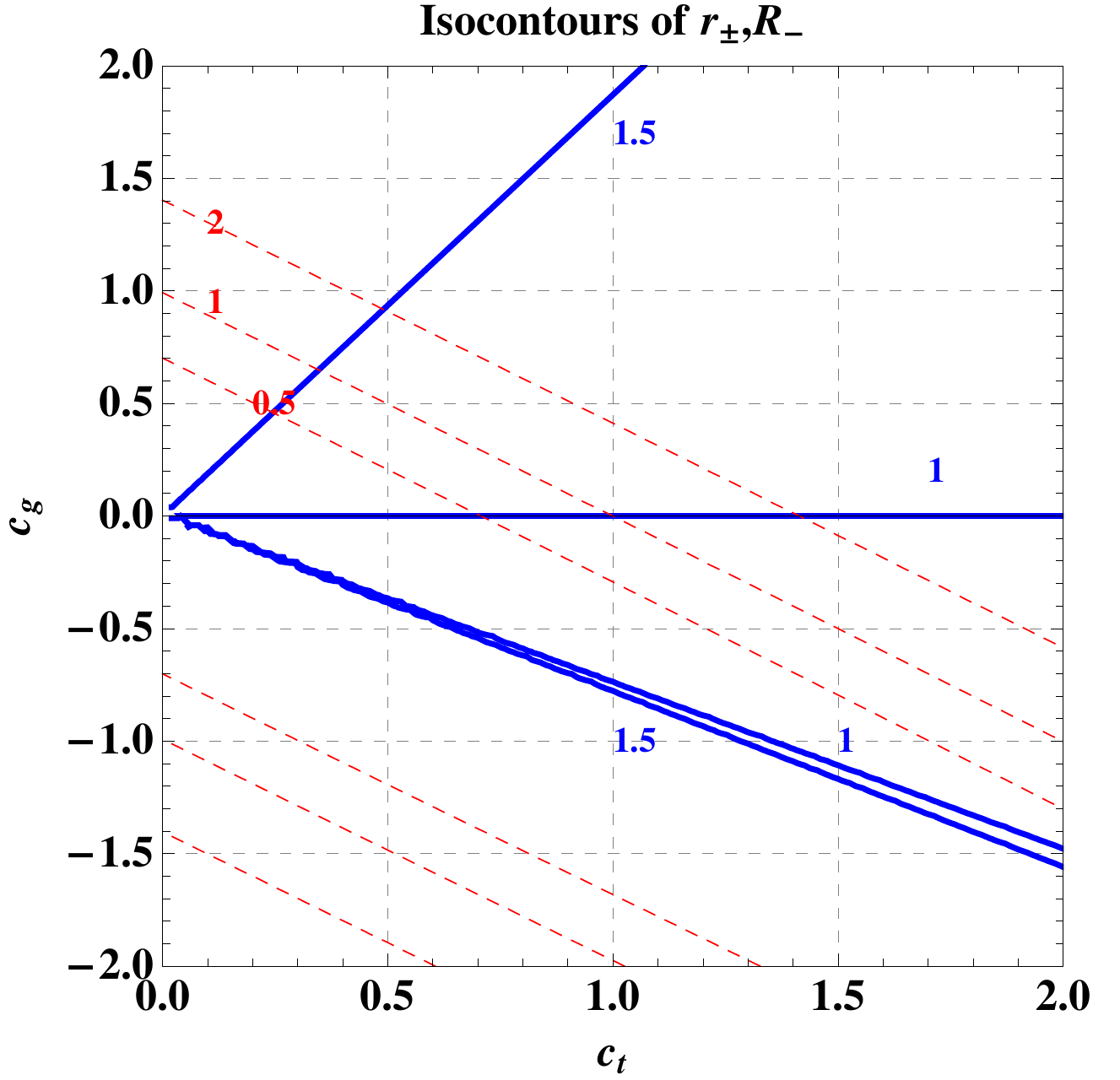}
\caption{$R_-$(red, dashed) and $r_\pm$(blue,solid) contours in the $(c_t,c_g)$ plane, $\sqrt{S}=14$TeV , $R_-$ and $r_\pm$ are defined for discriminating momentum $P_T=300$ GeV.}
\label{fig:rpmscat}
\end{center}
\end{figure}
 In the Fig. \ref{fig:rpmscat} we show the isocontours of $r_\pm$ in the $(c_t,c_g)$ plane. As one can see, there are significant parts of the $(c_t,c_g)$ space where one of $R_+$ or $R_-$ is diminished from unity while the other is enhanced heralding a presence of the heavy top partner. We have checked numerically that $r_{\pm}$ is almost independent of the PDF choice as well as the renormalization/factorization scale choice.  Also, note that the isocontours of the $r_\pm$ variable intercept with and are sometimes almost orthogonal to the $\frac{d\sigma }{d p_T}$ contours as can be seen from Fig. \ref{fig:rpmscat}, which illustrates that $r_{\pm}$ has good discriminating powers in the $(c_t, c_g)$ plane. 
 
 As an experimental observable $r_\pm$ can be expressed as
 \begin{equation}
 r_\pm=\frac{N^+/N^-}{\sigma^+_{SM}/\sigma^-_{SM}},
 \end{equation}
where $N^\pm$ is the number of events seen in respective bins. The denominator suffers from theoretical errors from the choice of scale and PDF errors but is independent of the $K$ factor and we have checked that the overall error is always 
$\lesssim 10\%$.
 This means that a LO estimate is sufficient for evaluating the denominator. The numerator suffers from experimental errors only. This ratio provides a definitive prescription for comparing an experimental measurement with a theoretical predictions with clearly delineated and disentangled experimental and theoretical errors.

%%%%%%%%%%%%%%
\section{Conclusion}
%%%%%%%%%%%%%%
We will conclude by recapitulating the main results of our work. The current data on the Higgs coupling shows a strong degeneracy in the best fit solutions for the Higgs couplings in the $(c_t, c_g)$ space.  
In this work we propose to use the $p p\ra h+j$   process to resolve this degeneracy. 
Indeed the Higgs interaction with gluons generated by the loops of the SM top quark and the dimension five operator have different $p_T$ dependence and this can be used to measure the effective Higgs couplings to tops and gluons.
To estimate the LHC potential  we have looked at the 4 lepton final state. Due to the very small rate of the signal this measurement can become possible only with very high luminosity at the LHC. The expected constraints on $c_t$ look, so far, inferior compared to the prospects in the direct measurements of the $t\bar{t}h$ coupling\cite{tthcoupling}(ATLAS projections for the $3000$ fb$^{-1}$ predict the determination of the top Yukawa coupling with a $\sim 10\%$ accuracy\cite{atlashl}). However $h+j$ can be used to reduce errors on the $c_g$ coupling when combined with the $c_t$ measurements from $t\bar{t}h$ production. Exploring other Higgs decay final states can also largely increase the precision of the $(c_t,c_g)$ measurements. 

We also propose an observables with reduced theoretical errors, $r_\pm$, which can be used as alternative discriminant of NP signal. We show that theoretical and experimental errors can be disentangled in $r_\pm$.

When this work was at the stage of completion we became aware of another project, which also uses $pp\ra h+j$ to measure Higgs couplings \cite{Grojean}.
%%%%%%%%%%%%%%%%%%%%%%%%%%%%%%%%%%%%%%%%%%%%%%%%%%%%%%
\section*{Acknowledgments}
 %%%%%%%%%%%%%%%%%%%%%%%%%%%%%%%%%%%%%%%%%%%%%%%%%%%%%%
%%%%%%%%%%%%%%%%%%%%%%%%%%%%%%%%%%%%%%%%%%%%%%%%%%%%%%
We would like to thank 
S. Rychkov and R. Contino for suggesting the project. We would also like to thank R. Contino for  discussions and encouragement throughout the completion of this work  and J. de Blas , M. Son and L. Silvestrini  for useful discussions. A. P. would like to thank the Physics Department of the University of Notre Dame du Lac for providing computational resources and hosting him over the summer of 2013 while this work was being done.
The work of A. A.  was supported by the ERC Advanced Grant No.~267985 
\textit{Electroweak Symmetry Breaking, Flavour and Dark Matter: One Solution for Three Mysteries (DaMeSyFla)}. The work of A. P. was supported by the European Research Council under the European Union's Seventh Framework Programme (FP/2007-2013) / ERC Grant Agreement n.~279972.

%%%%%%%%%%%%%%%%%%%%%%%%%%%%%%

%%%%%%%%%%%%%%%%%%%%%%%%%%%%%%%%5
\end{document}